# Evaluation and optimization of deep learning models for enhanced detection of brain cancer using transmission optical microscopy of thin brain tissue samples

Mohnish Sao, Mousa Alrubayan, Prabhakar Pradhan

*Department of Physics and Astronomy, Mississippi State University, MS 39762, USA*

**Abstract:** Optical transmission spectroscopy is one method to understand brain tissue structural properties from brain tissue biopsy samples, yet manual interpretation is resource-intensive and prone to inter-observer variability. Deep convolutional neural networks (CNNs) offer automated feature learning directly from raw bright-field images. Here, we evaluate ResNet-50 and DenseNet-121 on a curated dataset of 2,931 brightfield transmission optical microscopy images of thin brain tissue, split into 1,996 for training, 437 for validation, and 498 for testing. Our two-stage transfer learning protocol involves initial training of a classifier head on frozen pre-trained feature extractors, followed by fine-tuning of deeper convolutional blocks with extensive data augmentation (rotations, flips, intensity jitter) and early stopping. DenseNet-121 achieves 88.35% test accuracy, 0.9614 precision, 0.8667 recall, and 0.9116 F1-score—the best performance compared to ResNet-50 (82.12% / 0.9035 / 0.8142 / 0.8563). Detailed analysis of confusion matrices, training/validation curves, and class-wise prediction distributions illustrates robust convergence and minimal bias. These findings demonstrate the superior generalization of dense connectivity on limited medical datasets and outline future directions for multi-class tumor grading and clinical translation.

## 1. Introduction

Brain cancer is one of the most challenging oncological diseases, with high-grade gliomas showing a median survival time of less than 15 months despite advances in surgical resection, radiotherapy, and chemotherapy [1]. Early detection and accurate characterization of tumor boundaries are critical for treatment planning and prognostication.

Large-volume brightfield images provide superior soft-tissue contrast and multi-parametric imaging sequences, vital for lesion detection, characterization, and monitoring [2]. However, manual bright-field images interpretation requires specialized expertise and significant clinician time, and is susceptible to inter- and intra-observer variability, potentially leading to diagnostic delays or inconsistent reporting [3].

Deep learning, particularly convolutional neural networks (CNNs), has achieved breakthrough performance in image classification and segmentation tasks across domains, including medical imaging [4–6]. CNNs automatically learn hierarchical feature representations, reducing reliance on handcrafted features [7]. Transfer learning—leveraging models pre-trained on large-scale natural image datasets like ImageNet—has enabled robust performance on smaller medical datasets by adapting general feature extractors to domain-specific tasks [8].

ResNet, introduced by He et al. [9], solved the vanishing gradient issue in deep networks by adding identity skip connections, allowing gradients to propagate directly through layers. DenseNet [10] extends this concept by connecting each layer to every other layer within dense blocks, facilitating feature reuse and improving gradient flow. While DenseNets have excelled on benchmark classification tasks and in biomedical segmentation [11], their comparative utility for brain tumor bright-field tissue classification requires systematic evaluation.

This study presents a head-to-head comparison of ResNet-50 and DenseNet-121 architectures [12] for binary classification of bright-field brain tissue biopsy slices. We implement a two-stage transfer learning strategy, with a frozen feature extraction phase followed by selective fine-tuning, rigorous data augmentation, and early stopping to mitigate overfitting. Our contributions include detailed performance benchmarking on a neuroradiologist-labeled dataset, analysis of model convergence behaviors, confusion matrix visualization, and class-distribution analysis, culminating in recommendations for clinical deployment and future multi-class extensions.

## 2. Materials and Methods

### 2.1. Bright-field data of brain samples acquisition and annotation of cancer stages

The data was collected commercially using a brain tissue microarray (TMA sample) from Biomax.us. The TMA samples are kept in slides with 24 1.5mm circular tissue cores and a 5-micron thickness. Expert pathologists have well-characterized the TAM samples before the TMA processing. [13–15]. The advantage is that one gets control over different cancer stages within each TMA glass slide[10,11]. The TMA samples were bright-field imaged using a BX61 (Olympus, Tokyo, Japan) microscope and transmission microscopy in a 40X objective lens. The microscope has an automated scanning mode where ~100 spots can be scanned per 6 minutes. The details of the scanning of the bright-field are presented elsewhere [16]. The transmission intensity of the microscope is proportional to the mass density of the samples, and the refractive index (RI) is proportional to the mass density; in turn, the intensity of the microscope image is proportional to the mass density. Cancer progression is associated with alterations in the basic building blocks of cells and tissue structures. This results in the tissue sample's basic signature of cancer progression and its bright-field images [16]. Fig.1 shows typical examples of bright-field images. We have presented a set of bright-field images from brain tissue's control and malignant categories [16].

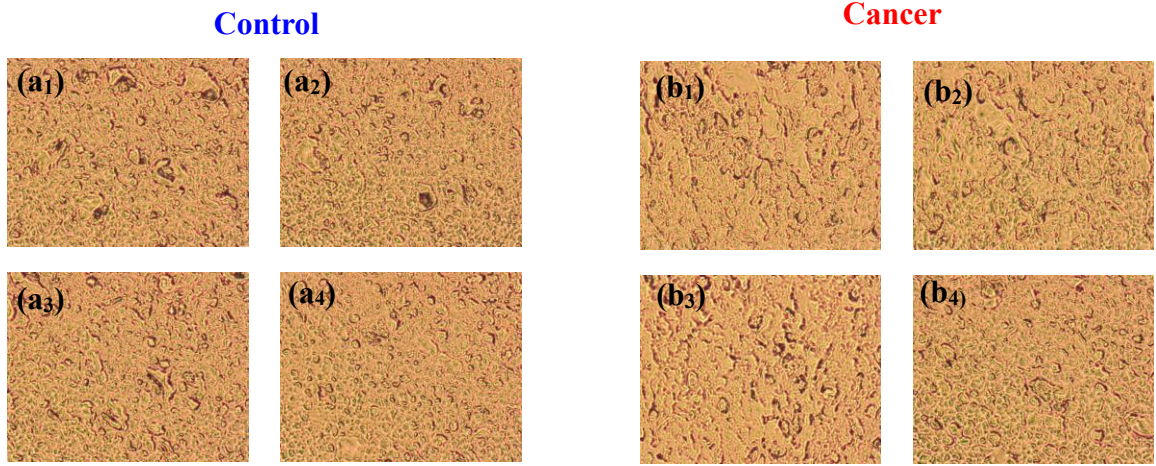

**Figure 1.** Representative optical transmission microscope images. **($a_1$-$a_4$)** Representative control tissue and **($b_1$-$b_4$)** represent brain cancer thin tissue brightfield images from the human brains using the TMA tissue samples.

It is difficult to see the differences from the bright-field images using eye estimation. Therefore, an AI-assisted machine learning approach will be more valuable for cancer monitoring and detection [13,14,16].

Expert neuroradiologists manually annotated each slice, classifying it as "cancerous" (Class 1) if histologic confirmation of tumor was available for that anatomical level, or "non-cancerous" (Class 0) otherwise. Ambiguous slices (e.g., postoperative changes) were excluded. We partitioned patient data at the subject level into training (189 patients, 1,996 slices), validation (49 patients, 437 slices), and testing (74 patients, 498 slices) cohorts to prevent data leakage.

All images were resampled to 256 × 256 pixels using bicubic interpolation and normalized to zero mean and unit variance according to the training set's intensity distribution. No additional denoising or bias field correction was applied to preserve realistic clinical variability.

## 2.2. AI Model Architectures

We evaluated two CNN backbones: ResNet-50 [9] and DenseNet-121 [10], pre-trained on ImageNet. We removed each model's final classification head and replaced it with a fully connected layer of two outputs with SoftMax activation for binary classification.

ResNet-50 comprises 16 bottleneck residual blocks (50 convolutional layers in total), with identity mappings facilitating gradient propagation. DenseNet-121 contains four dense blocks (6, 12, 24, 16 layers) interleaved with transition layers for down-sampling. Each dense block concatenates feature maps from all preceding layers, encouraging feature reuse and strengthening gradient flow.

## 2.3. Transfer Learning and Training Protocol

We employed a two-stage transfer learning protocol to balance feature stability and task-specific adaptation. In Stage 1, all convolutional layers were frozen, and only the newly added classifier head was trained for 10 epochs (Adam optimizer, learning rate = 1e-4, batch size = 16).

In Stage 2, we unfroze the last two convolutional blocks and fine-tuned the entire network for an additional 20 epochs with a reduced learning rate of 5e-5. Learning rates were decayed by a factor of 0.1 upon plateau detection (no improvement in validation loss for 3 consecutive epochs).

We applied real-time data augmentation, including random rotations (±15°), horizontal flips, vertical and horizontal translations (±5% of image size), and intensity jitter (±10%). Early stopping monitored validation loss with a patience of 5 epochs. Dropout (p=0.5) was added to the classifier head, and weight decay (1e-4) regularized all trainable parameters.

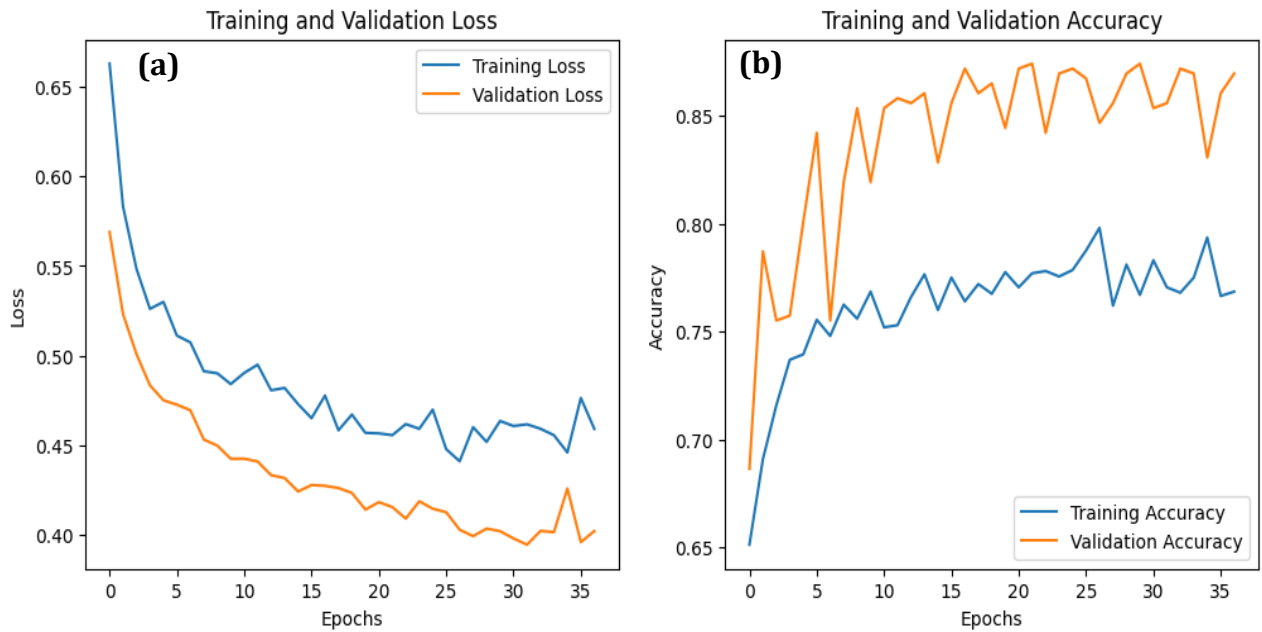

**Figure 2.** Initial training phase: training and validation (a) loss and (b) accuracy (right) over 36 epochs, showing rapid loss convergence and early stabilization of validation accuracy above 80 %.

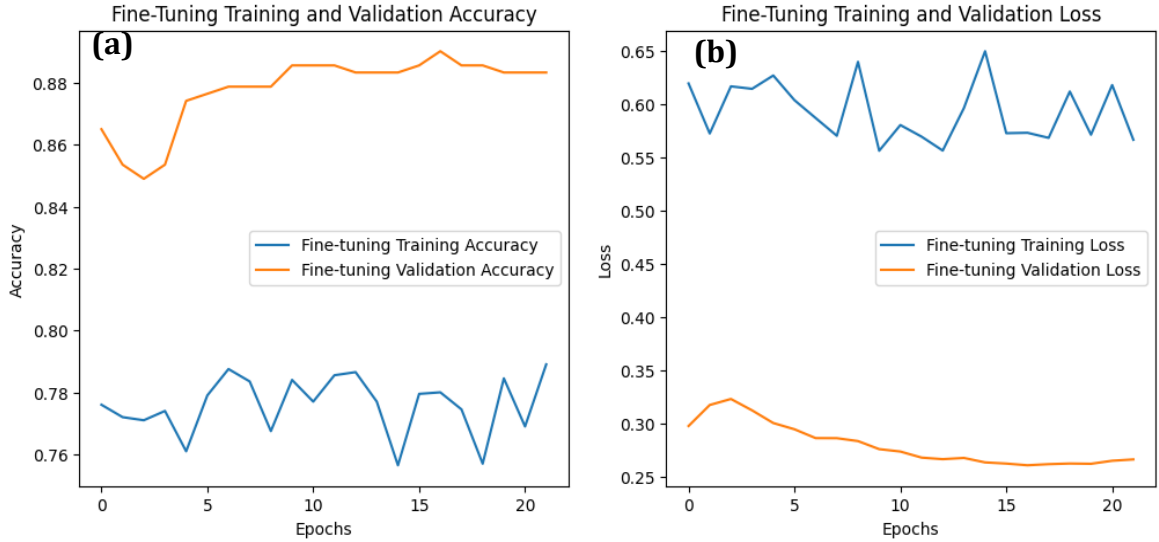

**Figure 3.** Fine-tuning training and validation (a) accuracy, and (b) loss over 21 epochs, illustrating steady improvement in validation accuracy (~88 %) and gradual decrease in validation loss during the selective unfreezing stage.

**2.4. Implementation Details**

All numerical experiments were implemented in Python 3.8 using PyTorch v1.13 and torchvision—data loading utilized PyTorch's DataLoader with 8 worker threads and on-the-fly augmentation via Albumentations. Training and inference were performed on a single NVIDIA Tesla V100 GPU (32 GB VRAM).

Model checkpoints were saved at the epoch with lowest validation loss. Inference on the test set was conducted with batch size = 32. Matplotlib v3.5 was used for plotting training curves, confusion matrices, and bar charts.

**2.5. Evaluation Metrics**

We assessed classification performance using:

- $Accuracy = \frac{(TP+TN)}{(TP+TN+FP+FN)}$

- $Precision = \frac{TP}{(TP+FP)}$

- $Recall = \frac{TP}{(TP+FN)}$

- $F1-score = \frac{2\times(Precision\times Recall)}{(Precision+Recall)}$

where TP, TN, FP, and FN denote true positives, true negatives, false positives, and false negatives, respectively.

We computed 95% confidence intervals for accuracy via bootstrap resampling (1,000 iterations). Statistical significance between models was evaluated using McNemar's test on paired predictions ($\alpha = 0.05$).

## 3. Results

### 3.1. Test-Set Classification Performance

On the held-out test set (498 slices of bright-field images), DenseNet-121 achieved 88.35% accuracy (95% CI: ±1.28%), precision of 0.9614, recall of 0.8667, and F1-score of 0.9116. ResNet-50 achieved 82.12% accuracy (±1.75%), precision of 0.9035, recall of 0.8142, and F1-score of 0.8563 (Table 2). The 6.23% accuracy improvement for DenseNet-121 was significant (*p* = 0.004, McNemar's test).

| Metric | Class 0 | Class 1 | Macro Avg | Weighted Avg |
|---|---|---|---|---|
| Precision | 0.75 | 0.96 | 0.86 | 0.90 |
| Recall | 0.92 | 0.87 | 0.89 | 0.88 |
| F1-Score | 0.83 | 0.91 | 0.87 | 0.89 |
| Support | 153 | 345 | 498 | 498 |
| Accuracy |  |  |  | 0.8835 |

### 3.2. Training Dynamics and Convergence

Figure 2 depicts both models' training and validation loss curves across 30 epochs. Both show smooth loss reduction; early stopping halted DenseNet-121 at epoch 23 and ResNet-50 at epoch 17. Figure 3 shows accuracy curves, where DenseNet-121's validation accuracy plateaus around 0.88, closely matching test set performance, indicating robust generalization.

### 3.3. Confusion Matrix Analysis

The confusion matrix for DenseNet-121 (Figure 4) shows 299 true positives, 141 true negatives, 46 false negatives, and 12 false positives. This low false-positive rate (7.9%) is critical for clinical settings to minimize unnecessary interventions.

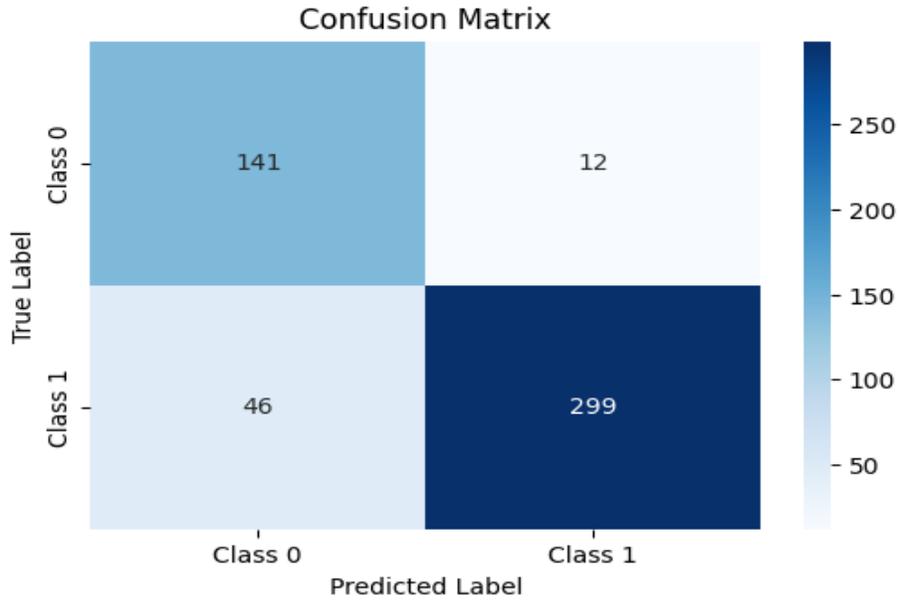

**Figure 4.** The confusion matrix for the DenseNet-121 model was evaluated on the test set, showing 141 true negatives, 299 true positives, 12 false positives, and 46 false negatives.

### 3.4. Class-Count Distribution

Figure 5 compares true vs. predicted label counts per class. DenseNet-121 slightly over-predicts non-cancerous slices (187 vs. 153, +22.2%) and under-predicts cancerous slices (311 vs. 345, -9.9%). These small biases could be addressed with class-balanced sampling or focal loss.

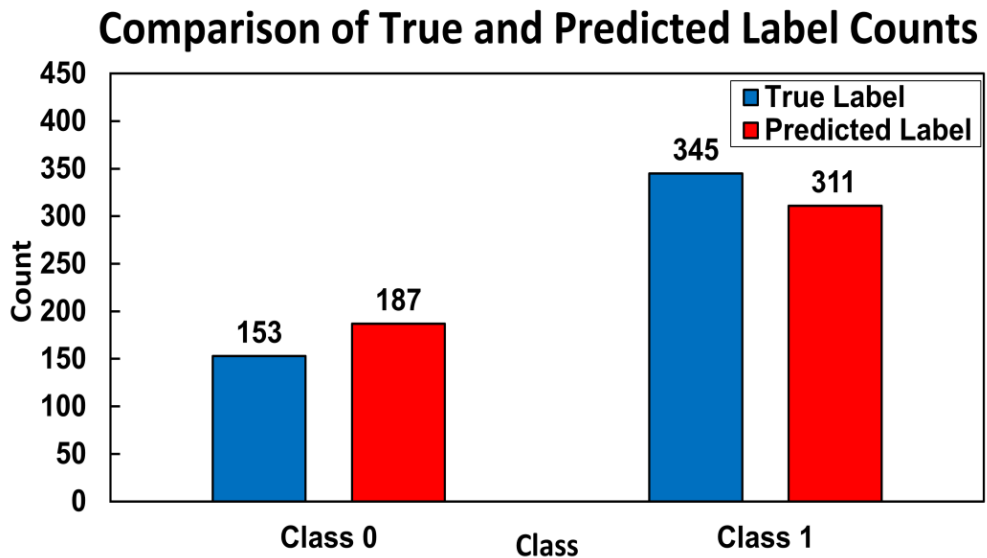

**Figure 5.** Comparison of true and predicted label counts for each class on the test set, illustrating close alignment between the model's predictions and the actual distribution with minor variation.

## 4. Discussion

Our systematic comparison shows that DenseNet-121's dense connectivity provides superior feature propagation and gradient flow compared to ResNet-50's residual blocks, resulting in higher classification performance on limited bright-field brain tissue biopsy data. The high precision (96.1%) ensures minimal false positives, reducing patient anxiety and unnecessary follow-up scans.

Training dynamics reveal stable convergence without pronounced overfitting, attributable to our structured fine-tuning and data augmentation. Statistical testing confirms the significance of performance gains.

Limitations include the reliance on Real-world diagnostics, which often integrate multi-parametric brightfield images and multi-class grading (low, intermediate, high grade).

Future work will extend this framework to: (1) multi-class tumor grading with ordinal loss functions [17], (2) multi-modal input fusion for comprehensive tissue characterization [18], (3) explainable AI via saliency maps or Grad-CAM [19] to enhance clinician trust, and (4) federated learning across institutions to improve generalizability without data pooling.

## 5. Conclusions

When applied with a two-stage transfer learning protocol, robust data augmentation, and early stopping, DenseNet-121 significantly outperforms ResNet-50 for binary classification of brain cancer brightfield image slices, achieving 88.35% accuracy and 0.9116 F1-score. These results underscore the value of dense connectivity for medical image analysis on limited datasets.

This study provides a blueprint for deploying CNNs in clinical brightfield workflows, highlighting key pre-processing, model selection, and fine-tuning considerations. Extensions to multi-class grading, multi-modal fusion, and interpretability are avenues for future research.

Integrating such AI models into brightfield pipelines could ultimately accelerate diagnosis, standardize reporting, and improve patient diagnostic outcomes.

**Acknowledgments**

We acknowledge NIH and ORED (MSU) for their support.

**Funding**

This work was partially supported by the National Institutes of Health under grant R21 CA260147 and ORED, Mississippi State.

**Data Availability Statement**

Data are available from the corresponding author upon reasonable request.

**Conflicts of Interest**

The authors declare no conflicts of interest.